\documentclass[10pt,journal,twocolumn]{IEEEtran}
\usepackage{times} 
\usepackage[cmex10]{amsmath}
\usepackage{bbm}
\usepackage{amssymb}
\usepackage{graphicx,dblfloatfix}
\usepackage{cite}
\usepackage{subfigure}
\usepackage{subfig}
\usepackage{subeqnarray}
\usepackage{verbatim}
\usepackage{xcolor}
\usepackage{enumerate}
\usepackage{stmaryrd}
\usepackage{multirow}
\usepackage{algorithm}
\usepackage{algpseudocode}
\usepackage{diagbox}
\usepackage{url}
\newcommand{\nc}{\newcommand}
\nc{\bsm}{\mbf}
\nc{\mbb}{\mathbb}
\nc{\mbs}{\mathbbmss}
\nc{\mbf}{\mathbf}
\nc{\mcl}{\mathcal}

\begin{document}

\title{Reconfigurable Intelligent Surfaces for EnergyEfficiency in D2D Communication Network}

\author{Shuaiqi Jia, Xiaojun Yuan, \IEEEmembership{Senior Member, IEEE}, and Ying-Chang Liang, \IEEEmembership{Fellow, IEEE}
\vspace{-.2cm}
\thanks{The authors are with the National Key Laboratory of Science and Technology on Communications and also with the Center for Intelligent Networking and Communications, University of Electronic Science and Technology of China, Chengdu 611731, China (e-mail: sqjia@std.uestc.edu.cn;xjyuan@uestc.edu.cn;liangyc@ieee.org).}
}

\maketitle

\begin{abstract}
In this letter, the joint power control of D2D users and the passive beamforming of reconfigurable intelligent surfaces (RIS) for a RIS-aided device-to-device (D2D) communication network is investigated to maximize energy efficiency. This non-convex optimization problem is divided into two subproblems, which are passive beamforming and power control. The two subproblems are optimized alternately. We first decouple the passive beamforming at RIS based on the Lagrangian dual transform. This problem is solved by using fractional programming. Then we optimize the power control by using the Dinkelbach method. By iteratively solving the two subproblems, we obtain a suboptimal solution for the joint optimization problem. Numerical results have verified the effectiveness of the proposed algorithm, which can significantly improve the energy efficiency of the D2D network.
\end{abstract}
\begin{IEEEkeywords}
D2D communication, power control, reconfigurable intelligent surface, passive beamforming. \vspace{-3mm}
\end{IEEEkeywords}

\section{Introduction}
\IEEEPARstart{E}{nergy} efficiency has emerged as a key performance indicator in the design of wireless communication systems, due to operational, economic, and environmental concerns \cite{buzzi2016survey}. Device-to-device (D2D) communication has been proposed as a promising technology to increase energy efficiency (EE), which allows new peer-to-peer communication service by leveraging the proximity and reuse gains \cite{jiang2015energy}. 
\par
Reconfigurable intelligent surface (RIS) is designed to improve the propagation environment and enhance wireless communications, and has attracted much attention recently \cite{8959174,8941126,huang2019reconfigurable}. Specifically, RIS is a meta-surface with a large number of passive reflecting elements. Each element can induce a phase shift to the incident signal independently. All elements cooperatively perform the reflect beamforming. Unlike the conventional amplify-and-forward (AF) relay, the RIS forwards the incident signals using passive reflection beamforming, which reduces the energy consumption of the system. These characteristics make the RIS technology attractive from an energy efficiency standpoint \cite{huang2019reconfigurable}.
\par
In this letter, we propose a RIS-assisted D2D communication network with a full-frequency reuse to support D2D users, which is assisted by some RISs. The EE maximization problem is formulated to optimize the RIS phase shifts and the transmit powers under phase-shift, maximum power and minimum transmission rate constraints. We divide the non-convex problem into two subproblems, which are passive beamforming and power control. The two subproblems can be optimized alternately. We first rewrite the passive beamforming at RIS based on the Lagrangian dual transform \cite{shen2018fractional}, and solve this subproblem based on the fractional programming \cite{shen2018fractional}. Then, we transform the power control subproblem in a fractional form into an equivalent optimization problem in a subtractive form, which is solvable by the Dinkelbach method \cite{dinkelbach1967nonlinear}. The two subproblems are solved iteratively until convergence, yielding a suboptimal solution for the joint optimization problem. Numerical results have verified the effectiveness of the proposed algorithm, which can significantly improve the energy efficiency of the D2D network.

\section{System Model}

In this letter, we consider the RIS-aided D2D communication network with full-frequency reuse to support D2D users. Without loss of generality, we assume that there are $L$ D2D links in the network, assisted by geographically separated $M$ RISs, where the $m$-th RIS, denoted by $\textrm{RIS}_{m}$, has $N_{m}$ elements. We assume that the power of the signals reflected by the RIS for two or more times can be ignored due to significant path loss. Besides, we assume that all the channels experience quasi-static flat-fading, and the channel state information (CSI) can be estimated by using the existing channel estimation methods, such as in \cite{8879620}. 
\par
\begin{figure}[h]
    \centering
    \includegraphics[scale=0.7]{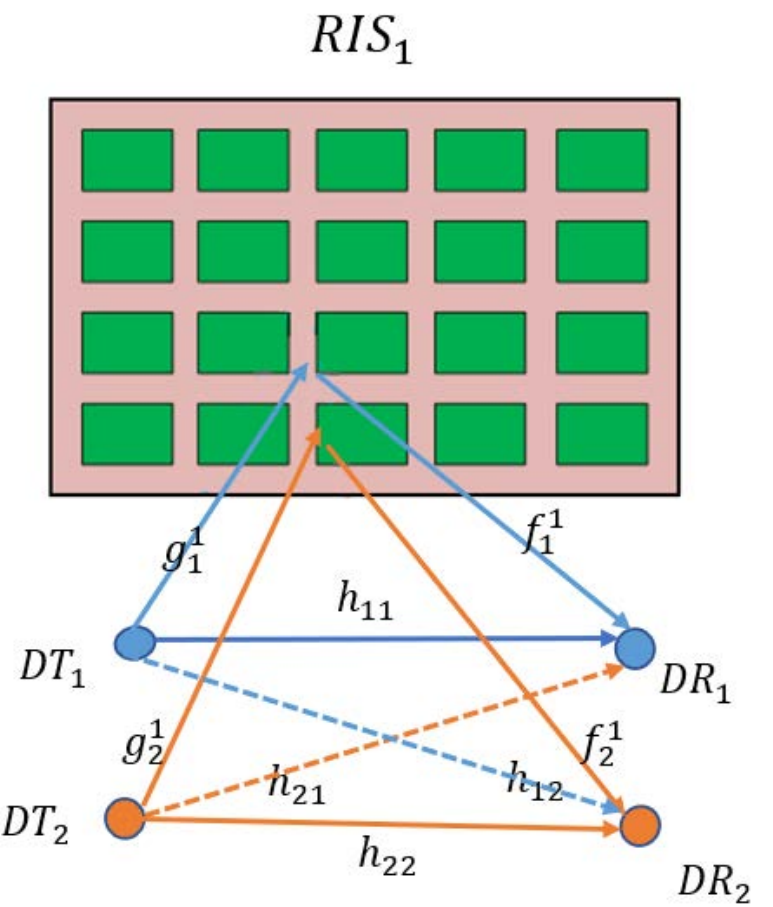}
    \caption{The RIS-aided D2D communication network with one RIS and two D2D pairs.}
    \label{f1}
\end{figure}

Denote by $h_{ij}\in \mathbb{C}$ the channel coefficient from transmitter $i$ to receiver $j$. The channel coefficient vectors from transmitter $i$ to $\textrm{RIS}_{m}$ and from receiver $j$ to $\textrm{RIS}_{m}$ are denoted by $\mbf{g}_{i}^{m}\in \mathbb{C}^{ N_{m}\times 1}$ and $\mbf{f}_{j}^{m}\in \mathbb{C}^{ N_{m}\times 1}$, respectively. Let $N=\sum_{m=1}^{M}N_m$ be the total number of reflecting elements. The channel coefficient vectors from transmitter $i$ to RISs and from receiever $j$ to RISs are denoted by $\mbf{g}_{i}\in \mathbb{C}^{ N\times 1}$ and $\mbf{f}_{j}\in \mathbb{C}^{ N\times 1}$, respectively. A simple scenario with one RIS and two D2D pairs is shown in Fig. \ref{f1}.
\par
The phase-shift matrix of the RISs is denoted by a diagonal matrix $\mbf{\Theta}=\sqrt{\eta} \operatorname{diag}\left(\boldsymbol{\theta}\right)$, where $\boldsymbol{\theta}=[\theta_{1}, \cdots, \theta_{N}]^{H}$, $|\theta_{n}|=1$ and $\eta \in (0,1)$. 
\par
Denote by $x_{i}$ the transmit data symbol to user $i$. The received signal of receiver $l$ is given by
\begin{equation}
    y_{l}=\sum_{i=1}^{L} (h_{il}+f_{l}^{T}\mathbf{\Theta}g_{i}) x_{i}+w_{l},
\end{equation}
where $w_{l} \sim \mathcal{C} \mathcal{N}\left(0, \sigma^{2}\right)$ is the additive white Gaussian noise (AWGN) at the $l$-th receiver. Denote by $p_{l}$ as the transmit power of D2D link $l$. The signal-to-interference-plus-noise ratio (SINR) of link $l$ is expressed as
\begin{equation}
z_{l}=\frac{|h_{l l}+ f_{l}^{T}\mathbf{\Theta}g_{l}|^{2}p_{l}} {\sum_{i \neq l,i=1}^{L} |h_{i l}+f_{l}^{T}\mathbf{\Theta}g_{i}|^{2} p_{i}+\sigma^{2}}.
\end{equation}
Then, the achievable sum rate of the system is given by
\begin{equation}
\mathcal{R} = \sum_{l=1}^{L} \log _{2}\left(1+z_{l}\right).
\end{equation}
\par
The total power consumption contains D2D users' transmit power $\mbf{p}$, circuit power $P_{c}$ and the RISs' power. The RISs' power consumption depends on the type and the resolution of its individual elements \cite{huang2019reconfigurable}. Elements with $b$-bit resolution can perform $b$-bit phase shifting on the impinging signal. The power consumption of the RISs with $N$ reflecting elements can be written as  
\begin{equation}
    P_{\mathrm{RIS}}=N P(b),
\end{equation}
where $P(b)$ denotes the power consumption of each element having $b$-bit resolution. Thus, the toal power consumption can be expressed as
\begin{equation}
\mathcal{P}_{\text {total }}=\sum_{l=1}^{L}\left( p_{l}+2P_{c}\right)+NP(b).
\end{equation}
\par
We define the ratio between the achievable sum rate and the total power consumption as the energy efficiency (EE). The energy efficiency is maximized by jointly optimizing the transmit power $\mbf{p}$ and the phase-shift matirx $\mbf{\Theta}$, subject to RIS’s phase-shift, maximum power and minimum transmission rate constraints. Thus, the problem can be formulated as
\begin{subequations} \label{P1}
\begin{align}
 \max_{\mbf{p}, \mbf{\Theta}} \quad & {\mathrm{EE}}=\frac{ \sum_{l=1}^{L} \log _{2}\left(1+z_{l}\right)}{\sum_{l=1}^{L} p_{l}+2L P_{c}+N P(b)} & \\
\mbox{s.t.}\quad
&\log _{2}\left(1+z_{l}\right) \geq R_{\min , l} \quad \forall l=1,2, \ldots, L  & \\
&0 \leq p_{l} \leq P_{\max } \quad \forall l=1,2, \ldots, L\\
& {\left|\theta_{n}\right|=1 \quad \forall n=1,2, \ldots, N}.&
\end{align}
\end{subequations}
It is difficult to obtain the optimal solution to this problem due to its nonconvex objective function and constant-modulus constraints. In the next section, an efficient algorithm is proposed to obtain a suboptimal solution to this problem.

\section{Energy Efficiency Maximization} \label{section3}
In this section, we adress the energy efficiency maximization problem of the RIS-aided D2D system. We divide the problem (6) into two subproblems, i.e., the passive beamforming and the power control, respectively. Then, the power control and passive beamforming can be optimized alternately.
\subsection{Optimizing $\boldsymbol{\Theta}$ for Given $\mbf{p}$}
For a fixed $\mbf{p}$, the problem (\ref{P1}) reduces to
\begin{align} \label{P2}
\max_{ \boldsymbol{\Theta}} \quad &~  \sum_{l=1}^{L} \log _{2}\left(1+z_{l}\right) \notag \\
\mbox{s.t.}\quad &(6b),(6d).
\end{align}
\par
We apply the the Lagrangian dual transform \cite{guo2019weighted} to (\ref{P2}), with the new objective function given by
\begin{equation}
\begin{aligned}
f(\boldsymbol{\Theta},\boldsymbol{\beta})=\sum_{l=1}^{L} \log _{2}&~\left(1+\beta_{l}\right) -\sum_{l=1}^{L} \beta_{l} \\
&+\sum_{l=1}^{L} \frac{(1+\beta_{l})|h_{l l}+ f_{l}^{T}\mathbf{\Theta}g_{l}|^{2}p_{l}}{\sum_{i=1}^{L} |h_{i l}+f_{l}^{T}\mathbf{\Theta}g_{i}|^{2} p_{i}+\sigma^{2}},
\end{aligned}
\end{equation}
where $\boldsymbol{\beta}$ refers to a set of auxiliary variables. We propose to optimize $\boldsymbol{\beta}$ and $\boldsymbol{\Theta}$ alternately. For a fixed $\boldsymbol{\Theta}$, the optimal $\beta_{l}$ is
\begin{equation}
    \beta_{l}^{\circ}=\frac{|h_{l l}+ f_{l}^{T}\mathbf{\Theta}g_{l}|^{2}p_{l}} {\sum_{i \neq l,i=1}^{L} |h_{i l}+f_{l}^{T}\mathbf{\Theta}g_{i}|^{2} p_{i}+\sigma^{2}}.
\end{equation}
Then, for a fixed $\boldsymbol{\beta}$, optimizing $\boldsymbol{\Theta}$ is reduced to
\begin{align} \label{P3}
\max_{ \boldsymbol{\Theta}} \quad &  f_{1}(\boldsymbol{\Theta})=\sum_{l=1}^{L}\frac{(1+\beta_{l})|h_{l l}+ f_{l}^{T}\mathbf{\Theta}g_{l}|^{2}p_{l}}{\sum_{i=1}^{L} |h_{i l}+f_{l}^{T}\mathbf{\Theta}g_{i}|^{2} p_{i}+\sigma^{2}}& \notag \\
\mbox{s.t.}\quad
&(6b),(6d).&
\end{align}

\par
The problem (\ref{P3}) can be solved by using the FP method. Define $\mathbf{A}_{il}=diag(f_{l})g_{i}\sqrt{p_{i}}$ and $b_{il}=h_{il}\sqrt{p_{i}}$. Using the quadratic transform proposed in \cite{shen2018fractional}, $f_{1}(\boldsymbol{\Theta})$ is reformulated as 
\begin{equation} \label{obj1}
\begin{aligned}
f_{2}(\boldsymbol{\theta},\boldsymbol{\varepsilon})=\sum_{l=1}^{L} 2\sqrt{1+\beta_{l}}&~\mathbf{Re}\{ 
\varepsilon_{l}^{*}\boldsymbol{\theta^{\mathbf{H}}}\mathbf{A}_{ll}+\varepsilon_{l}^{*}b_{ll}
\}\\
&-\sum_{l=1}^{L}|\varepsilon_{l}|^{2} (\sum_{i=1}^{L}|b_{il}+\boldsymbol{\theta^{\mathbf{H}}}\mathbf{A}_{il}|^{2}+\sigma^{2}).
\end{aligned}
\end{equation}
\par
We optimize $\boldsymbol{\theta}$ and $\boldsymbol{\varepsilon}$ alternately. The optimal $\varepsilon_{l}$ can be obtained by letting $\partial f_{2}/ \partial \varepsilon_{l}=0$, yielding
\begin{equation}
    \varepsilon_{l}^{\circ}=\frac{\sqrt{1+\beta_{l}}(b_{ll}+\boldsymbol{\theta^{\mathbf{H}}}\mathbf{A}_{ll})}{\sum_{i=1}^{L}|b_{il}+\boldsymbol{\theta^{\mathbf{H}}}\mathbf{A}_{il}|^{2}+\sigma^{2}}.
\end{equation}
\par
Then the remaining problem is to optimize $\boldsymbol{\theta}$ for a given $\boldsymbol{\varepsilon}$. By simplifying (\ref{obj1}), the optimization problem for $\boldsymbol{\theta}$ is represented as
\begin{align} \label{P4}
 \max_{ \boldsymbol{\theta}} \quad &  f_{3}(\boldsymbol{\theta})= -\boldsymbol{\theta}^\mathbf{H}\mathbf{U}\boldsymbol{\theta}+2\textbf{Re}\{\boldsymbol{\theta}^\mathbf{H}\mbf{v} \}+C \notag \\
\mbox{s.t.}\quad
&(6b),(6d),&
\end{align}

where
\begin{equation}
\mathbf{U}=\sum_{l=1}^{l}\left|\varepsilon_{l}\right|^{2} \sum_{i=1}^{L} \mathbf{A}_{il} \mathbf{A}_{il}^{\mathrm{H}}
\end{equation}
\begin{equation}
\mbf{v}=\sum_{l=1}^{L}\left(\sqrt{\beta_{l}+1} \varepsilon_{l}^{*} \mathbf{A}_{ll}-\left|\varepsilon_{l}\right|^{2} \sum_{i=1}^{l} b_{il}^{*} \mathbf{A}_{il}\right),
\end{equation}
and $C$ is a constant. Since $\mbf{A}_{il}\mbf{A}_{il}^{\mathrm{H}}$ for all $i$ and $l$ are positive-definite matrices, $\mathbf{U}$ is a positive-definite matrix and $f_{3}(\boldsymbol{\theta})$ is a quadratic concave functions of $\boldsymbol{\theta}$. Problem (\ref{P4}) is non-convex and inhomogeneous due to the non-convexity of the quadratic constraints and the unit modulus constraints. Letting $\boldsymbol{\bar{\theta}^{\mbf{H}}}=[\boldsymbol{\theta^\mbf{H}}, 1]$ and $\mbf{Q}=\boldsymbol{\bar{\theta}}\boldsymbol{\bar{\theta}^\mbf{H}}$, and dropping the rank-one constraint of $\mathbf{Q}$, problem (\ref{P4}) can be rewritten as
\begin{subequations} \label{P5}
\begin{align}
\max_{ \mathbf{Q}} \quad &  \operatorname{Tr}(\mathbf{\bar{U}} \mathbf{Q}) & \\
\mbox{s.t.}\quad
&\gamma_{l}^{\min }\left(\sum_{i \neq l}^{L}\left(\operatorname{Tr}\left(\mbf{R}_{i, l} \mathbf{Q}\right)+\left|b_{il}\right|^{2}\right)+\sigma^{2}\right) &\notag\\
&\leq \operatorname{Tr}\left(\mbf{R}_{l, l} \mathbf{Q}\right)+\left|b_{ll}\right|^{2}, \forall l, l=1, \ldots, L& \\
& \mathbf{Q}_{n, n}=1, \forall n=1, \ldots, N+1&\\
&\mathbf{Q} \succcurlyeq 0,&
\end{align}
\end{subequations}
where,
\par
$\mbf{R}_{i, l}=\left[\begin{array}{cc}
\mbf{A}_{il} \mbf{A}_{il}^{\mathrm{H}} & \mbf{A}_{il} b_{il}^{*} \\
b_{il} \mbf{A}_{il}^{\mathrm{H}} & 0
\end{array}\right]$, $\mathbf{\bar{U}}=\left[\begin{array}{cc}
 \mathbf{U}& \mbf{-v}\\
\mbf{-v}^{\mathrm{H}} & 0
\end{array}\right]$.
\par
We can solve problem (\ref{P5}) via CVX \cite{cvx}. Then the standard Gaussian randomization can be used to obtain a feasible rank-one solution. We summarize the optimization method for $\mathbf{\Theta}$ with a fixed $\mbf{p}$ in Algorithm 1.
\begin{algorithm}[H]
\vspace{0.2mm}
\caption{\textbf{\!:} \vspace{0.15mm} Optimizing $\boldsymbol{\Theta}$ for Given $\mbf{p}$ }
\textbf{\,Initialization:} Initialize $\boldsymbol{\Theta}^{(0)}$ to a feasible value, $i=0$.
\begin{algorithmic}[1]
\State Update the auxiliary variable $\beta_{l}^{i}$ by (9).
\\ Update the auxiliary variable $\epsilon_{l}^{i}$ by (12).
\\Update $\boldsymbol{\Theta}^{(i)}$ by solving (\ref{P5}) together with Gaussian randomization.
\\Repeat setps 1-3 until the value of $\boldsymbol{\Theta}$ converges.
\end{algorithmic}
\textbf{\,Output:} $\boldsymbol{\Theta}^{(i)}$
\end{algorithm}

\subsection{Optimizing $\mbf{p}$ for Given $\boldsymbol{\Theta}$}
For a fixed $\boldsymbol{\theta}$, the problem (\ref{P1}) reduces to
\begin{align} \label{P6}
\max_{\mbf{p}} \quad&\frac{ \sum_{l=1}^{L} \log _{2}\left(1+z_{l}\right)}{\sum_{l=1}^{L} p_{l}+2L P_{c}+N P(b)} & \notag \\
\mbox{s.t.}\quad
&(6b),(6c).& 
\end{align}
Problem (\ref{P6}) is a non-concave FP problem. From \cite{dinkelbach1967nonlinear}, we define a new optimization problem as
\begin{align} \label{P7}
\max_{\mbf{p}} \quad&~F(\lambda) & \notag \\
\mbox{s.t.}\quad
&(6b),(6c),& 
\end{align}
where $\lambda$ is a non-negative parameter, and
\begin{equation} \label{obj2}
  F(\lambda)={\sum_{l=1}^{L} \log _{2}\left(1+z_{l}\right)}-\lambda({\sum_{l=1}^{L} p_{l}+2L P_{c}+N P(b)}). 
\end{equation}
$F(\lambda)$ is continuous and strictly monotonically decreasing in $\lambda$ and has a unique root $\lambda^{*}$. The optimal solution $\mbf{p^{*}}$ of problem (\ref{P6}) is the same as that of problem (\ref{P7}) with $\lambda=\lambda^{*}$, where $\lambda^{*}$ can be obtained by using the Dinkelbach method \cite{dinkelbach1967nonlinear}.
\par
The key of the method is to solve problem (\ref{P7}) for a given $\lambda$. To be specific, (\ref{obj2}) can be rewritten as
\begin{equation}
    F(\lambda)=f_{1}(\mbf{p})-f_{2}(\mbf{p})
\end{equation}
where
\begin{equation}
    \begin{aligned}
    f_{1}(\mbf{p})=\sum_{l=1}^{L}\log_{2}(\sum_{i=1}^{L}&~|h_{il}+f_{l}\boldsymbol{\Theta}g_{i}|^{2}p_{i}+\sigma^{2})\\
    &-\lambda \sum_{i=1}^{L}p_{i}-\lambda(2LP_{c}+NP(b))
    \end{aligned}
\end{equation}
\begin{equation}
    f_{2}(\mbf{p})=\sum_{l=1}^{L}\log_{2}(\sum_{i=1,i\neq l}^{L}|h_{il}+f_{l}\boldsymbol{\Theta}g_{i}|^{2}p_{i}+\sigma^{2}).
\end{equation}
Clearly, $f_{1}(\mbf{p})$ and $f_{2}(\mbf{p})$ are concave. Hence, the objective function of (\ref{P7}) is the difference of two concave functions. Constraints in (6b) are also the difference of two concave functions, which are
\begin{equation}
    c_{1,l}(\mbf{p})-c_{2,l}(\mbf{p}) \geq R_{\min , l} \quad \forall l=1,2, \ldots, L 
\end{equation}
where
\begin{align}
   c_{1,l}(\mbf{p})=\log_{2}(\sum_{i=1}^{L}|h_{il}+f_{l}\boldsymbol{\Theta}g_{i}|^{2}p_{i}+\sigma^{2})
\end{align}
\begin{align}
     c_{2,l}(\mbf{p})=\log_{2}(\sum_{i=1,i\neq l}^{L}|h_{il}+f_{l}\boldsymbol{\Theta}g_{i}|^{2}p_{i}+\sigma^{2}).
\end{align}
\par
Consequently, problem (\ref{P7}) is a difference of convex (DC) programming problem. We apply the DC algorithm \cite{tao1997convex} to obtain a suboptimal solution of problem (18). 

\section{Numerical Result}
In this section, numerical results are provided to demonstrate the validity of the proposed algorithm. We consider a RIS-aided D2D network with multiple single-antenna D2D users, which are assumed randomly and uniformly placed in the $200m \times 200m$ rectangular. And there are four RISs in the rectangular. The distance between D2D users is 20-40m. The baseband channels of the RIS-user link, user-RIS link, and user-user link are modeled as Rician fading channels with Rician factor $\beta =2$. The background noise at the receivers is $\sigma^{2}=-117$ dBm. We set the path-loss constant $k=10^{-3}$, the path-loss exponent $\chi=4$. Circuit power at each user $P_{C}=15$ dBm. RIS reflection efficiency $\eta=0.8$. All presented results are obtained by averaging over 1000 independent channel realizations.
\par 
We compare the performance of the proposed algorithm with two baselines. Baseline 1 is the energy efficiency optimized by power control without the aid of RIS. Baseline 2 is the energy efficiency optimized by optimal power control with the aid of RISs. We adopt the Branch and Bound method to obtain the optimal power control \cite{yang2016energy}. Fig. 2 illustrates the energy efficiency of different power parameters $P(b)$ with respect to the size of $N$ of RIS. As we can see, significant performance gains are achieved by joint power control and RISs' passive beamforming optimization. Typical power consumption values of each RIS element are 1.5, 4.5, 6, and 7.8 mW for 3-, 4-, 5-, and 6-bit resolution \cite{ribeiro2018energy}. As the number of quantized bits increases, the energy efficiency decreases gradually. Although high-bit RIS will improve the spectral efficiency of the network, its high power consumption leads to the decrease of energy efficiency.
\begin{figure}[h]
    \centering
    \includegraphics[scale=0.5]{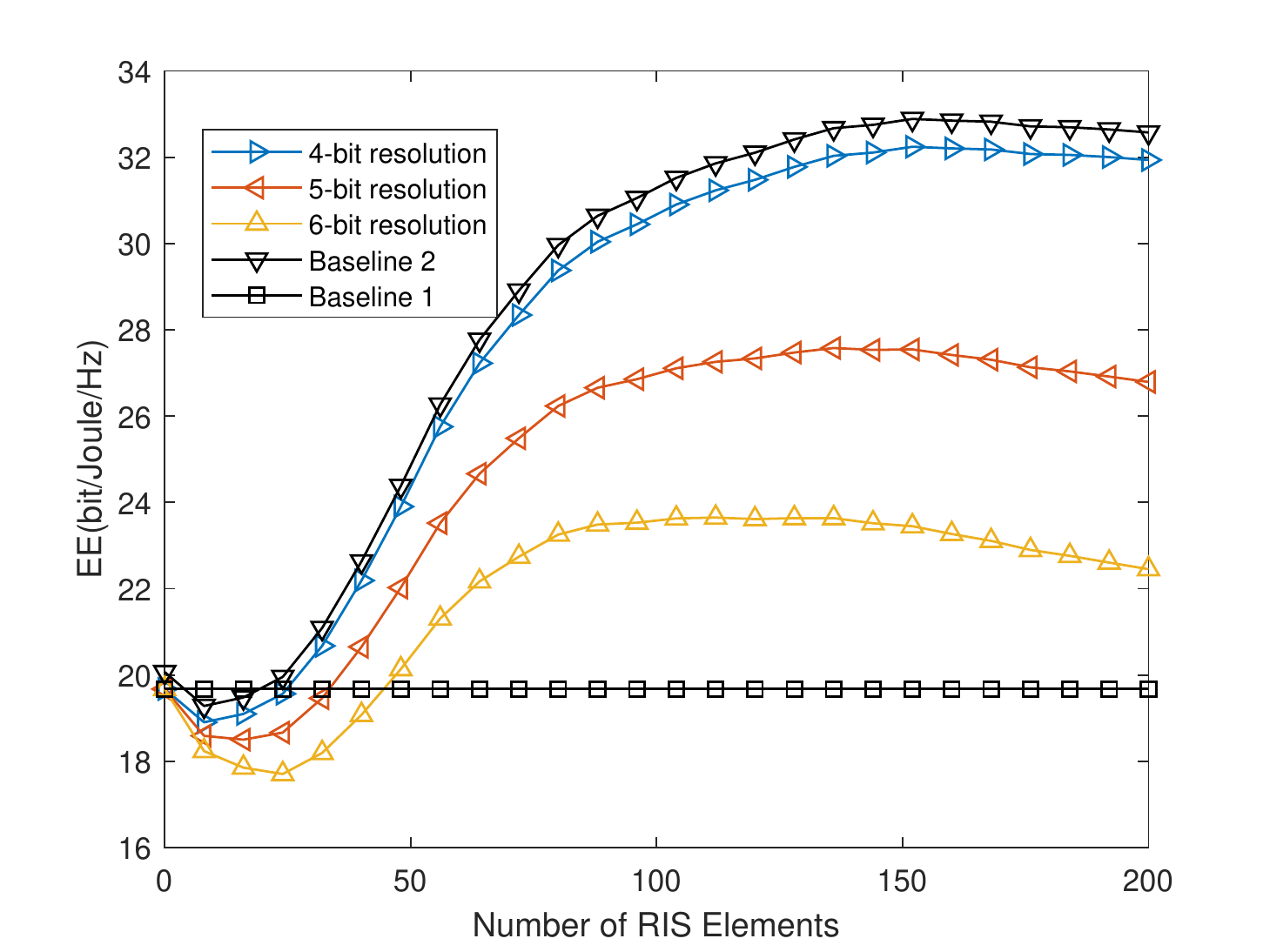}
    \caption{The energy efficiency versus the number of RIS elements for  $L=10$, $P_{max}=0.1$ W, $R_{min}=0$ bps/Hz}
    \label{4}
\end{figure}
\par
Particularly, EE performance increases as $N$ increases. However, for a large $N$, EE starts decreasing. This is because for a large $N$, RISs' improvement in spectrum efficiency is no longer enough to compensate for the decrease in system energy efficiency caused by high power consumption. So, there exists an optimal number $N$ of RIS elements. Besides, the energy efficiency tends to decrease firstly and then increase. This is because when $N$ is too small, there will be little improvement in spectral efficiency with the assistance of the RISs. However, the power consumption of the RISs results in the decrease of energy efficiency.

\begin{figure}[t]
   \centering
    \includegraphics[width=4.5cm,height=4.5cm]{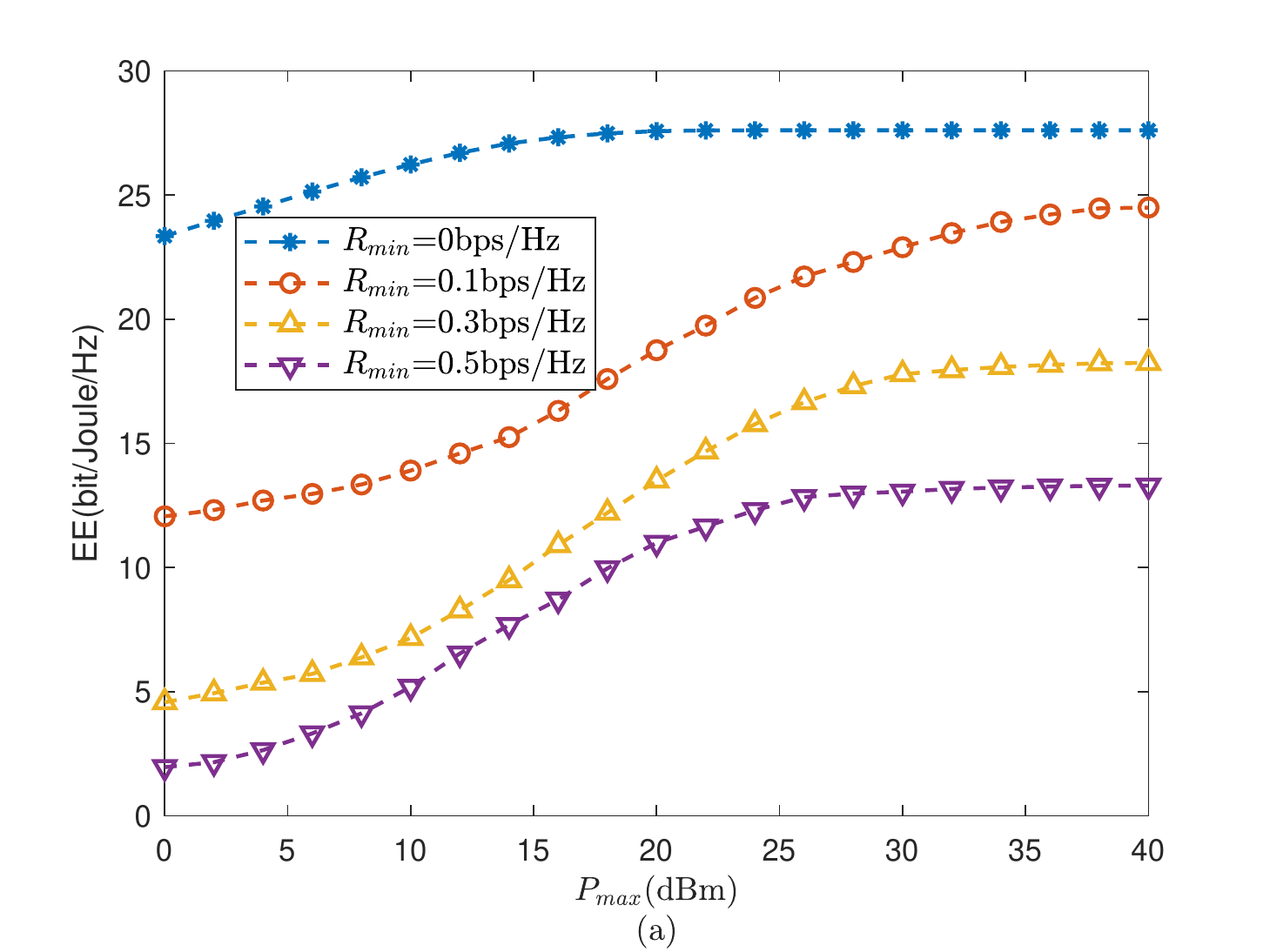}\,\includegraphics[width=4.5cm,height=4.5cm]{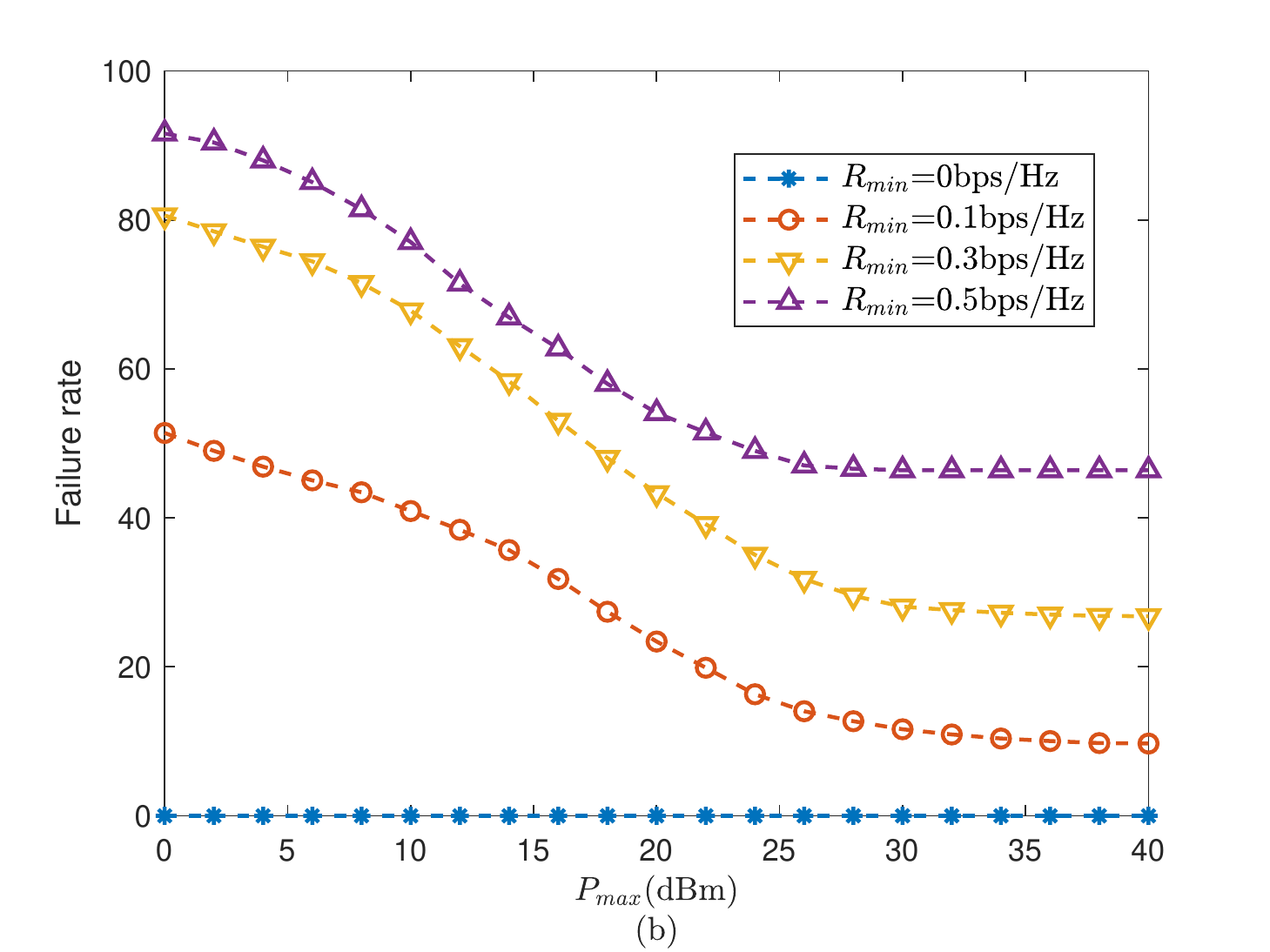}\\
    \vspace{-2.5mm}
    \caption{(a) The energy efficiency versus $P_{max}$ for $L=10$, $N=80$; (b) The failure rate versus $P_{max}$ for $L=10$, $N=80$.}
    \label{fig1}
\end{figure}
\par
The effect of the different values of $R_{min}$ and EE performances versus $P_{max}$ in dBm is depicted in Fig. 3. For the cases where the design problems turned out to be infeasible, EE is set to zero, corresponding to a communication failure. Fig. 3(a) compares the EE with the different minimum rate $R_{min}$. When $R_{min}=0$, EE increases to a fixed value as the power limit $P_{max}$ increases. For the higher $R_{min}$, EE will increase to a lower fixed value. Fig. 3(b) compares the failure rate with the different minimum rate $R_{min}$. Similarly, for the higher $R_{min}$, the failure rate will reduce to a higher fixed value. At this moment, increasing $P_{max}$ does not improve efficiency or reduce failure rate.
\section{Conclusion}
In this letter, the RIS technique was applied to enhance the energy efficiency of D2D communication network. To maximize the energy efficiency, we proposed a joint power control and RISs' passive beamforming optimization algorithm for obtaining the high-quality suboptimal solution. Simulation results demonstrated that the assistance of the RISs is beneficial to substantially improve the energy efficiency of the D2D communication network.

\renewcommand\refname{References}
\bibliographystyle{IEEEtran}
\bibliography{bib-he}

\end{document}